\documentclass{ringb99}
\usepackage{graphics}
\usepackage{amssymb}
\begin{document}
\title{Relations of Physical Properties for a Large Sample of X-Ray
Galaxy Clusters} 
\author{Thomas H. Reiprich \and Hans B\"ohringer}  
\institute{Max-Planck-Institut f\"ur extraterrestrische Physik,
 Giessenbachstra{\ss}e 1, 85740 Garching, Germany}
\maketitle

\begin{abstract}
One hundred and six galaxy clusters have been studied in
detail using X-ray data and important global physical properties have been
determined. Correlations between these properties are studied with
respect to the description of clusters as a self-similar class of
objects. The results support the self-similar picture only partly and
significant deviations are found for some correlations.
\end{abstract}

\section{Introduction}
Galaxy clusters are the largest collapsed objects in the
Universe. Understanding the physical processes governing the formation
of these enormous objects therefore represents an intriguing
task. Relations between bulk properties of clusters provide an
understanding of 
the underlying physics. There is now high quality X-ray data
available for many clusters and recently many authors have studied
cluster samples in order to search for such fundamental relations
(e.\,g., Fukazawa 1997; Allen \&
Fabian 1998 (also these proceedings, atp); Arnaud \&
Evrard 1999; Ettori \& Fabian 1999 (atp); Horner et al.\ 1999; Jones
\& Forman 1999; Mohr et al.\ 1999; Schindler 1999 (atp)).

We have constructed an X-ray flux-limited sample of the brightest
clusters in the ROSAT All-Sky Survey (RASS) to determine the local
cluster mass function (Reiprich 1998, Reiprich \& B\"ohringer 1999a). 
This sample as well as some additionally included clusters not meeting
the strict selection
criteria represent a great opportunity to correlate the physical
properties of nearby clusters. Here we give the preliminary results.
Throughout $H_0=h_{50}^{-1}\,50\,\rm km/s/Mpc$, $h_{50}=1$,
$q_0=0.5$ and $\Lambda=0$ is used. 

\section{Data Reduction and Analysis}
We used mainly high exposure ROSAT PSPC pointed observations to determine the
surface brightness profiles of the clusters, excluding obvious point sources.
If no pointed PSPC observations were
available in the archive or if clusters were too large for the field of view
of the PSPC we used RASS data. To calculate the gas density profiles
the standard $\beta$-model (Cavaliere \& Fusco-Femiano 1976),
(Gorenstein et al.\ 1978) (Equ.\ \ref{REIPRICH_beta1}) has been
used.
\begin{equation}
\rho_{\rm gas}(r)=\rho_{\rm gas}(0)\left(1+\frac{r^{2}}{r_{\rm c}^{2}}\right)^{-\frac{3}{2}\beta}
\label{REIPRICH_beta1}
\end{equation}
\begin{equation}
S_{\rm X}(R)=S_{\rm X}(0)\left(1+\frac{R^{2}}{r_{\rm c}^{2}}\right)^{-3\beta+\frac{1}{2}}
\label{REIPRICH_beta2}
\end{equation}
Fitting the corresponding surface brightness formula (Equ.\ \ref{REIPRICH_beta2})
to the observed surface brightness profiles gives the parameters needed to derive
the gas density profiles. To check if the often detected central
excess emission (central surface brightness of a cluster exceeding the fit
value) biases the mass determination we also fitted a double $\beta$-model of the
form $S_{\rm X}=S_{\rm X_1}+S_{\rm X_2}$ and calculated the gas mass profiles by
$\rho_{\rm gas}=\sqrt{\rho_{\rm gas_1}^2+\rho_{\rm gas_2}^2}$. Comparison of
the single and double $\beta$-model gas masses shows good agreement.

We compiled the values for the gas temperature ($T_{\rm gas}$) from the literature,
giving preference to
temperatures measured by the ASCA satellite (Markevitch et al.\ 1998;
Fukazawa 1997; Edge \& Stewart 1991; David et al.\ 1993).
For clusters for which we did not find a published temperature we used
the X-ray luminosity-temperature relation given by Markevitch (1998).

Assuming hydrostatic equilibrium the gravitational masses for the clusters can
be determined. Using the ideal gas equation, plugging Equ.\
\ref{REIPRICH_beta1} into the hydrostatic equation and
assuming the intracluster gas to be isothermal yields the gravitational mass
profile
\begin{equation}
M_{\rm tot}(r)=\frac{3kT_{\rm gas}r^{3}\beta}{\mu m_{\rm
p}G}\left(\frac{1}{r^2+r_{\rm c}^2}\right).
\label{REIPRICH_hy}
\end{equation}

Having acquired the gravitational mass profiles for the clusters it is now
important to determine the radius within which to determine the cluster mass.
Simulations by Evrard et al.\ (1996) have shown that the assumption of hydrostatic
equilibrium is generally valid within a radius, $r_{500}$, where the mean gravitational
mass density is greater than or equal to 500 times the critical density
$\rho_{\rm c}=4.7\cdot10^{-30}\rm g\,cm^{-3}$, as long as clusters undergoing
strong merger events are excluded. We calculated
$M_{\rm tot}$ at $r_{500}$ and also $r_{200}$ which is usually referred
to as the virial radius. Using these definitions of the outer radius instead of
a fixed
length, e.\,g., one Abell radius ($3\,h_{50}^{-1}\,\rm Mpc$), also
allows the uniform treatment of clusters of different size.
Using $r_{500}$ also saves us from an extrapolation much beyond the significantly
measured cluster emission in general.

Simulations by Schindler (1996) and Evrard et al.\ (1996) have shown
that determined and true cluster mass do not differ dramatically ($\lesssim
20\,\%$)
if clusters undergoing strong merger events are excluded.

There
are currently contradictory measurements for average radial $T_{\rm
gas}$ gradients (e.\,g., Markevitch et al.\ 1998, Irwin et al.\ 1999,
Kikuchi et al.\ 1999), if $T_{\rm gas}$ decreases with increasing $r$
on average, as Markevitch et al.\ have shown,
the isothermal assumption leads to
a systematic overestimation of $\sim 30\,\%$ of $M_{\rm tot}$ at $\sim
6$ core radii.

\section{Results}
\label{REIPRICH_res}
\begin{figure}
 \resizebox{\hsize}{!}{\includegraphics{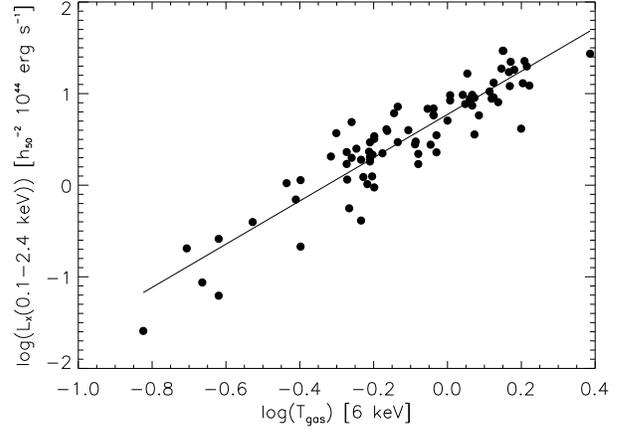}}
\caption[]{$L_{\rm X}-T_{\rm gas}$ relation.}\label{REIPRICH_fig1}
\end{figure}

\begin{figure}
 \resizebox{\hsize}{!}{\includegraphics{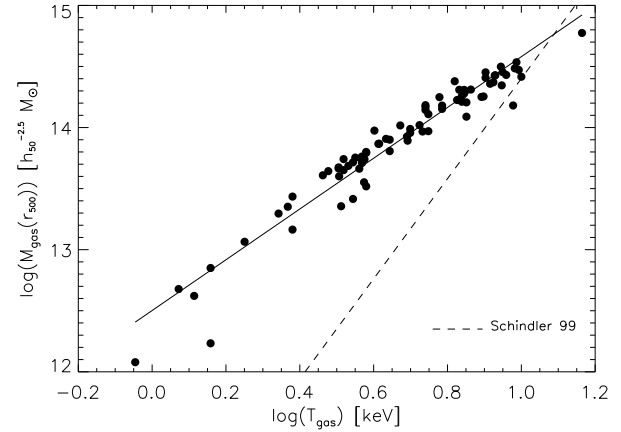}}
\caption[]{$M_{\rm gas}-T_{\rm gas}$ relation.}\label{REIPRICH_fig5}
\end{figure}

\begin{figure}
 \resizebox{\hsize}{!}{\includegraphics{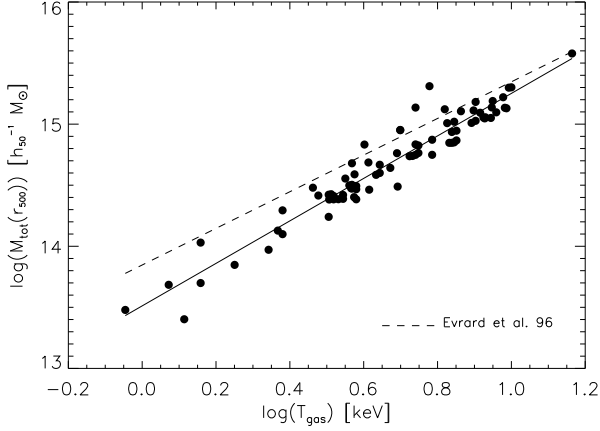}}
\caption[]{$M_{\rm tot}-T_{\rm gas}$ relation.}\label{REIPRICH_fig2}
\end{figure}

\begin{figure}
 \resizebox{\hsize}{!}{\includegraphics{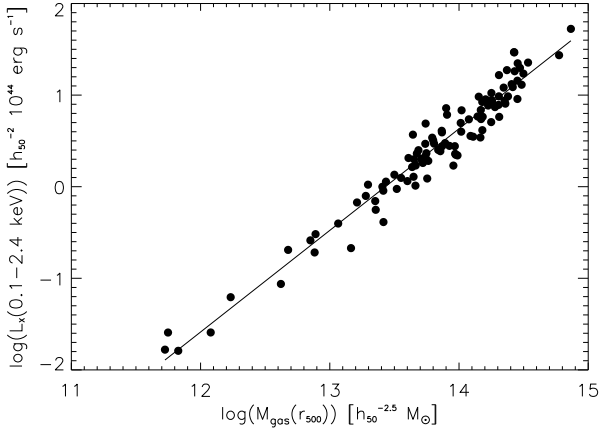}}
\caption[]{$L_{\rm X}-M_{\rm gas}$ relation.}\label{REIPRICH_fig6}
\end{figure}

\begin{figure}
 \resizebox{\hsize}{!}{\includegraphics{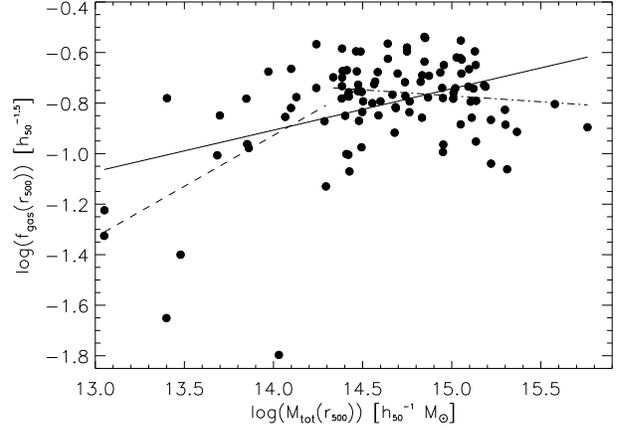}}
\caption[]{$f_{\rm gas}-M_{\rm tot}$ relation. Solid line: best fit 106
clusters; dashed line: best fit 20 clusters
$<2.0\cdot10^{14}M_{\odot}$; dot-dashed line: best fit 86 clusters
$\ge2.0\cdot10^{14}M_{\odot}$.}\label{REIPRICH_fig3}
\end{figure}

\begin{figure}
 \resizebox{\hsize}{!}{\includegraphics{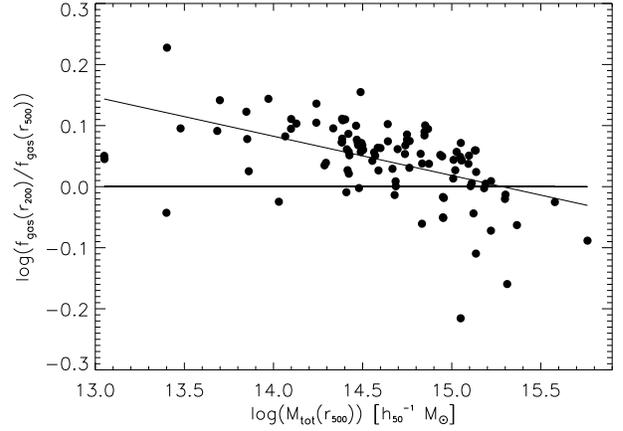}}
\caption[]{$F_{2/5}-M_{\rm tot}$ relation.}\label{REIPRICH_fig4}
\end{figure}

For correlations of the gas temperature with other quantities we
exclude the 24 clusters whose temperatures have been estimated using
the $L_{\rm X}-T_{\rm gas}$ relation. Luminosities have been calculated in
the energy range $0.1-2.4\rm\, keV$.
In Fig.\ \ref{REIPRICH_fig1} the $L_{\rm X}-T_{\rm gas}$
relation for 82 clusters is shown. A linear regression fit to the
logarithmically 
plottet data points yields $L_{\rm X}=5.9\cdot 10^{44}\,T_{\rm
gas}^{2.36}$, where $L_{\rm X}$ is in units of erg s$^{-1}$ and $T_{\rm
gas}$ in 6 keV. There is an indication of a steepening of the slope
towards the lower temperatures.
%
In Fig.\ \ref{REIPRICH_fig5} the $M_{\rm gas}-T_{\rm gas}$
relation for 82 clusters is shown. A linear regression fit
yields $M_{\rm gas}=3.2\cdot 10^{12}\,T_{\rm gas}^{2.08}$, where
$M_{\rm gas}$ is in solar units and $T_{\rm gas}$ in keV. Also here a
steeper slope for low $T_{\rm gas}$-clusters is visible. The slope is
less steep and the normalization much higher compared to the high
redshift, high temperature sample of Schindler (1999).
%
In Fig.\ \ref{REIPRICH_fig2} the $M_{\rm tot}-T_{\rm gas}$
relation for 82 clusters is shown. A linear regression fit yields
$M_{\rm tot}=3.2\cdot 10^{13}\,T_{\rm
gas}^{1.74}$, where $M_{\rm tot}$ is in solar units and $T_{\rm
gas}$ in keV. The slope is steeper and the normalization lower
compared to the simulated clusters of Evrard et al.\ (1996).
%
In Fig.\ \ref{REIPRICH_fig6} the $L_{\rm X}-M_{\rm gas}$
relation for 106 clusters is shown. A linear regression fit yields
$L_{\rm X}=1.3\cdot 10^{29}\,M_{\rm gas}^{1.11}$, where $L_{\rm X}$
is in units of erg s$^{-1}$ and $M_{\rm gas}$ in solar units. There is
a subtle hint that high $M_{\rm gas}$-clusters exhibit a steeper slope
than ones with low $M_{\rm gas}$.
%
In Fig.\ \ref{REIPRICH_fig3} the $f_{\rm gas}-M_{\rm tot}$
relation for 106 clusters is shown. A linear regression fit yields
$f_{\rm gas}=6.2\cdot 10^{-4}\,M_{\rm tot}^{0.16}$, where $f_{\rm
gas}$ ($=M_{\rm gas}/M_{\rm tot}$, gas mass fraction) is
unitless and $M_{\rm tot}$ in solar units.
The slope of this relation depends on the relative
number of lightweight clusters included, as can be seen from the fit
of two power laws ($f_{\rm gas_1}\propto M_{\rm tot_1}^{0.40}$,
$f_{\rm gas_2}\propto M_{\rm tot_2}^{-0.05}$) for different mass ranges.
%
In order to asses the radial
distribution of the gas mass fractions compared to the mass
of the clusters we have also determined the gas and gravitational
masses within $r_{200}$. If the ratio $F_{2/5}=f_{\rm
gas}(r_{200})/f_{\rm 
gas}(r_{500})$ is greater than one, then the radial gas distribution
is less steep than the dark matter distribution in the outer parts of
the clusters.
In Fig.\ \ref{REIPRICH_fig4}
the $F_{2/5}-M_{\rm tot}$ relation for 106 clusters is shown. A
linear regression fit yields $F_{2/5}=
9.6\cdot \,M_{\rm tot}^{-0.06}$, where $F_{2/5}$ is unitless and
$M_{\rm tot}$ in solar units. The slope
gets steeper with increasing $M_{\rm tot}$.

\section{Discussion}
In the previous section clear, tight correlations have been shown to
exist between global physical parameters of galaxy clusters.
How do these correlations compare to theoretical predictions and
relations found by other authors? What can we infer from the measured
relations?

Simple self-similar scaling laws predict $T_{\rm gas}\propto M_{\rm
tot}^{1.5}$ and $L_{\rm X}\propto f_{\rm gas}^2 M_{\rm tot}$ at a
characteristic radius, e.\,g. $r_{500}$, in the ROSAT energy band ($0.1
- 2.4\rm\, keV$). In this self-similar picture $f_{\rm gas}(M_{\rm
tot},M_{\rm gas})$ should be
constant for all clusters.
We find, however, $f_{\rm gas}\propto M_{\rm tot}^{0.16}$ (Fig.\
\ref{REIPRICH_fig3}) and 
$f_{\rm gas}\propto M_{\rm gas}^{0.22}$ (graph not shown). We note
the closer similarity to a
broken power law shape, given in section \ref{REIPRICH_res}.
The
following derived simple theoretical relations often depend on $f_{\rm
gas}$, therefore we give both, the values derived with constant $f_{\rm
gas}$ and the values derived with the found dependencies of $f_{\rm
gas}$ on $M_{\rm tot},M_{\rm gas}$ (in parentheses).

One derives $L_{\rm X}\propto M_{\rm tot}^{1.0\,(1.32)}$, with 1.32
being consistent with 
the empirical $L_{\rm X}-M_{\rm tot}$ relation we have determined
($L_{\rm X}=3.3\cdot 10^{26}\, M_{\rm tot}^{1.23}$; Reiprich
\& B\"ohringer 1999b). With the above proportionalities  
one finds $L_{\rm X}\propto T_{\rm gas}^{1.5\,(1.98)}$, not very well in
agreement with our measured result $L_{\rm X}\propto T_{\rm
gas}^{2.36}$ (Fig.\ \ref{REIPRICH_fig1}). Markevitch (1998) found
$L_{\rm X}\propto T_{\rm
gas}^{2.09}$. In his sample, however, no low temperature clusters are
included ($T_{\rm gas} \ge 3.5\rm\, keV$),
which seem to steepen the relation slightly in our sample.

For the $M_{\rm
gas}-T_{\rm gas}$ relation one would expect a proportionality like 
$M_{\rm gas} \propto T_{\rm gas}^{1.5\,(1.74)}$, which does not compare well
with our measured relation $M_{\rm gas} \propto T_{\rm
gas}^{2.08}$ (Fig.\ \ref{REIPRICH_fig5}). The gas mass is determined almost independently from the
temperature, so this tight correlation indicates small intrinsic
scatter. Our measurement is in agreement with Mohr et al.\ (1999) who
found $M_{\rm gas} \propto T_{\rm
gas}^{1.98}$. Schindler (1999) found for a sample of high redshift,
high temperature clusters an exponent of $4.1\pm 1.5$ and a factor of
$\sim 100$ lower
normalization. These large differences may indicate an
evolutionary effect, in the sense that for $T_{\rm gas}\lesssim 10\rm\,
keV$ high redshift clusters have a higher temperature at the same gas
mass. This interpretation is supported by the fact that Schindlers
$L_{\rm X,bol} - T_{\rm gas}$ relation has a steeper slope than the
same relation for nearby samples. But the uncertainties are too large
for a final decision.

Our measured relation $M_{\rm tot} \propto T_{\rm gas}^{1.74}$ (Fig.\
\ref{REIPRICH_fig2}) is
steeper than the expected value of 1.5 for the exponent. A similar
discrepancy has also been found by other authors (Mohr et al.\ 1999;
Horner et al. 1999,
using data of Fukazawa; Ettori \& Fabian 1999), however,
Horner et al.\  show for a sample of medium to high mass clusters that the discrepancy
may be reconciled if measured $T_{\rm gas}$ profiles are used to
determine $M_{\rm tot}$ (they used $r_{200}$). The normalization of the $M_{\rm tot}
- T_{\rm gas}$ relation of the simulated clusters of Evrard
et al.\ (1996) is a factor of $\sim 2$ higher than our measured
normalization. In general one should keep in mind that $T_{\rm gas}$ and
$M_{\rm tot}$ are not two independently measured quantities (Equ.\
\ref{REIPRICH_hy}), which complicates the interpretation of the
fitting result. 

Scaling laws predict $L_{\rm X}\propto M_{\rm gas}^{1.0\,(1.22)}$.
We measure $L_{\rm X}\propto M_{\rm gas}^{1.11}$ (Fig.\ \ref{REIPRICH_fig6}),
which is in reasonable agreement.

If clusters were exactly self-similar objects, the gas mass fraction at a
characteristic radius should be constant for all clusters. Several
authors have related measured values of $f_{\rm gas}$ and $M_{\rm
tot}$ or $T_{\rm gas}$. Up to now there is not even qualitative
agreement on the form of this relation. For example David et al.\
(1995) found indications for an increase of $f_{\rm gas}$ with
increasing $T_{\rm gas}$, Allen \& Fabian (1998) for a sample of
X-ray luminous clusters found indications for a decrease of $f_{\rm gas}$ with
increasing $T_{\rm gas}$, Ettori \& Fabian (1999) found no dependence of
$f_{\rm gas}$ on $M_{\rm tot}$ for high luminosity clusters, but
instead found indications for a decrease of $f_{\rm gas}$ with increasing
redshift. We find $f_{\rm gas} \propto M_{\rm tot}^{0.16}$ (Fig.\
\ref{REIPRICH_fig3}) but
as noted earlier a simple power law description of this relation may
not be very useful. The gas mass fraction stays fairly constant for $M_{\rm
tot}\gtrsim 10^{14} M_{\odot}$ but seems to decrease strongly below
this mass.

We next consider the gas mass fractions determined within two different
characteristic radii ($r_{500}$ and $r_{200}$) and relate the ratio
of these two fractions to the gravitational mass. Fig.\
\ref{REIPRICH_fig4} exhibits several interesting features. A) The gas
mass fractions are in general larger when measured within larger
radii, as noted by previous authors (e.\,g., David et al.\ 1995). This
means the gas distribution is flatter than the dark matter
distribution in the outer parts of clusters. B) The increase of
$f_{\rm gas}$ with radius gets smaller towards higher mass clusters. A
qualitatively similar trend has also been found by Schindler (1999). C) For
clusters with $M_{\rm tot}\gtrsim 10^{15} M_{\odot}$ the gas mass fractions
seem to decrease with increasing radius. 
These features are qualitatively consistent with non-gravitational
energy input affecting less massive groups/clusters more than
massive clusters (e.\,g., supernova driven galactic winds). These
features do not suggest a universal value
for the gas mass fraction, $f_{\rm gas}$ rather
seems to depend on
radius and cluster mass. Feature C) indicates the possibility that
estimates of $\Omega$ based on extrapolation of
$f_{\rm gas}$ within a
specific cluster radius to the whole Universe may deserve
reconsideration. For
instance if the trend of decreasing gas mass fraction towards higher
masses and therefore larger radii continues, $\Omega$ may have been
underestimated previously. But: it is not yet clear if one can draw significant
conclusions from Fig.\ \ref{REIPRICH_fig4} since determination of
masses within $r_{200}$ generally requires extrapolation
beyond the significantly measured cluster emission. Furthermore the
hydrostatic assumption, fundamental for the
gravitational mass estimate, may not always be justified at $\sim
r_{200}$.
Therefore this relation should be considered as tentative. We are
currently in the process of estimating the errors -- which can be large
since the gravitational mass plays a key role -- in order to quantify
the significance of the results.

\vspace{-0.08cm}
\section{Conclusions}
1.) We find several tight correlations between physical parameters of a
large sample of galaxy clusters.\\
2.) Not all of the slopes agree with expectations from simple self-similar
scaling relations and some also cannot be made compatible if variations
in the gas mass fractions are taken into account.\\
3.) Slight deviations from pure power laws are found. Except the weak bend in
the $L_{\rm X} - M_{\rm gas}$ relation they
may qualitatively be explained by assuming that a non-gravitational
energy input has taken place in clusters.\\
4.) Comparison to relations of high redshift clusters gives a weak
indication for 
evolution, in the sense that clusters at high redshift have a higher
gas temperature at a given gas mass.\\
5.) It is suggested that it may be difficult to assign a
universal gas mass 
fraction to galaxy clusters, since the gas mass fraction seems to depend on
radius and cluster mass.\\
6.) For massive clusters indications are found
that the gas mass fraction decreases in the outer parts. If confirmed
this may influence estimates of the
baryon density in the Universe based on extrapolations from
gas mass fractions on cluster scale to Universe scale.

\vspace{-0.08cm}

\end{document}